\documentclass[entropy,perspective,accept,pdftex,moreauthors]{Definitions/mdpi}

\usepackage{csquotes}
\firstpage{1}
\pubvolume{1}
\issuenum{1}
\articlenumber{0}
\pubyear{2023}
\copyrightyear{2023}
\externaleditor{Academic Editors: Irena Vodenska, Carolina Mattsson, Paolo Barucca and Giulio Cimini}
\datereceived{10 April 2023}
\daterevised{31 August 2023} 
\dateaccepted{7 September 2023}
\datepublished{ }
\hreflink{https://doi.org/} 

\Title{An Introduction to Complex Networks in Climate Finance}

\TitleCitation{An Introduction to Complex Networks in Climate Finance}

\Author{Alexander P. Kartun-Giles 
 * and Nadia Ameli }

\AuthorNames{Alexander P. Kartun-Giles and Nadia Ameli }

\AuthorCitation{Kartun-Giles, A.P.; Ameli, N.}

\address[1]{Institute for Sustainable Resources, The Bartlett School of Environment, Energy and Resources, University College London, London, WC1H 0NN, 
 UK; n.ameli@ucl.ac.uk }

\corres{\hangafter=1 \hangindent=1.05em \hspace{-0.82em}Correspondence: alexander.giles@ucl.ac.uk}

\abstract{In this perspective, we introduce recent research into the structure and function of complex investor networks supporting sustainability efforts. Using the case of solar, wind and hydro energy technologies, this perspective explores the complexity in low-carbon finance markets, defined as markets that direct capital flows towards low-carbon technologies, using network approaches to study their structure and dynamics. Investors are modeled as nodes which form a network or higher-order network connected by edges representing projects in which joint funding or security-related insurance was provided or other investment-related interaction occurred. We review the literature on investor networks generally, particularly in the case of complex networks, and address areas where these ideas were applied in this emerging field. The complex investor dynamics which emerge from the extant funding scenarios are not well understood. These dynamics have the potential to result in interesting non-linear behaviour, growth, and decline, which can be studied, explained and controlled using the tools of network science. }


\begin{document}


\section{Introduction}\label{sec:1}

Financial markets are complex systems~\cite{ladyman2012,mantegna1999}. When it comes to the study of the financial system for climate action, meaning capital flows directed towards low-carbon interventions with direct greenhouse gas mitigation benefits, complexity is a paradigm shift. The~financial system, as~any other complex system, reflects the interactions and dynamics of its heterogeneous actors across the globe who have diverse investment preferences and operate in different market conditions, which collectively shape the actual flows and investments in low carbon~\cite{hall2015,schweitzer2009}. Complexity is in both its composition (heterogeneous agents interacting simultaneously with each other on multiple levels, e.g.,~national and international) and in the diversity of emergent behaviour~\cite{foster2005,an2009}. Analyses incorporating such aspects are vital for policy design to monitor and influence the way financial markets could pool long-term financial assets to boost the low-carbon~transition.

\begin{figure}[H]
   
\begin{adjustwidth}{-\extralength}{0cm}
   \includegraphics[width=6.8in]{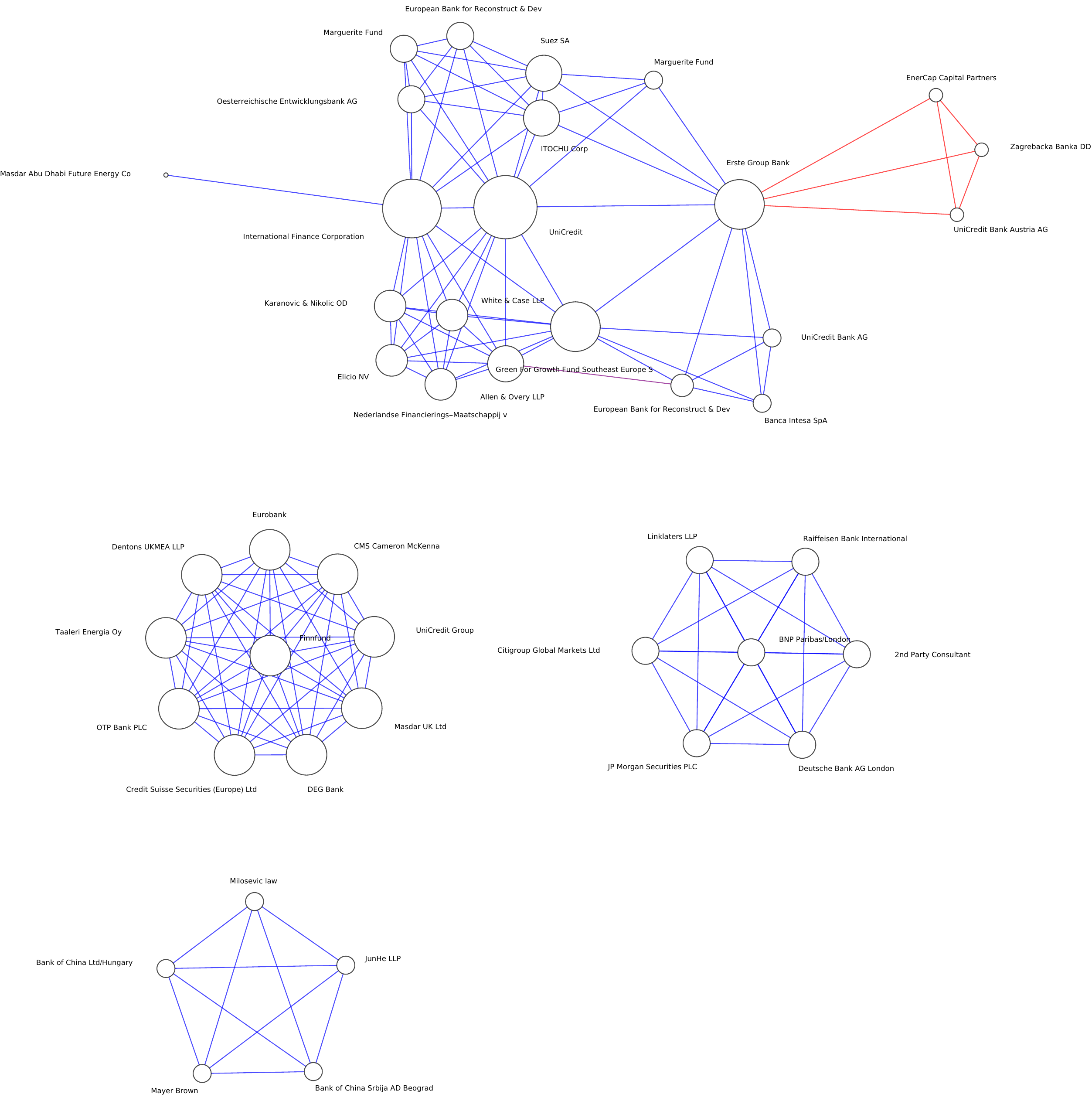}
\end{adjustwidth}
   \caption{Green 
 debt issued in the Balkans to date. Nodes are banks, and~links exist between banks when they insure a bond issuance together (i.e., investors provide financing for a loan to support a green energy project, and~the loan is insured by the larger financial system by buying the debt and reselling to investors as a security). A~multilayer network is formed, since banks work together on deals that are domiciled in a specific country. Whenever two banks work together to underwrite a loan for a project whose country of domicile is listed in Kosovo, they are connected with a blue link, with~a red link when in Bosnia, and~with a purple link in Montenegro. The~node degree is reflected in its relative size. The~Austrian financial service provider \textit{{Erste Group Bank} 
} has \textit{{activity}} 2, since it takes part in two layers, and~\textit{{degree}} 11, since it has interacted with that many~banks.}
 \label{fig:1}
\end{figure}

Research focuses on how techniques from theoretical physics can be used to study economic systems of this type~\cite{ameli2021,larosa2022,rickman2022, barthelemy2011, grimmettbook}. As~a system of interacting human agents, economic actors form a much more sophisticated complex system than that of the interacting spins in a disordered magnetic state such as a spin glass, but~the principles are the same, such as what Galton called The Law of Frequency of Error: ``The huger the mob, and~the greater the apparent anarchy, the~more perfect is its sway'' \cite{pearson2011}. As~with flocks of starlings forming a murmuration, we also have the collective behaviour of banking syndicates funding low-carbon projects, see Fig. \ref{fig:1}, and~despite acting  independently based on strict financial self-interest, significant emergent behaviour arises that can be understood both mathematically and economically, and~for societal~benefit. 
 
Moreover, this behaviour can be universal, in~that large classes of systems all with varying microscopic differences can have the same behaviour in a certain respect~\cite{barthelemy2011,kartungiles20192}. A~famous example is in theoretical ecology, where Robert May shows how  (models of) food-webs are stable (in terms of population numbers) below a universal critical interaction strength, independent of the probability distribution from which interaction strengths are drawn~\cite{may1972}. In~recent years, however, we turn to the grandest scale, and~look at how these ideas apply in  real systems such as political groups, social networks, city formations, and~beyond. It is the aspect of \textit{{randomness}} emerging from \textit{{deterministic}} laws in these systems that unites them under the theme of complexity, and~it can be remarkable to see case studies of probabilistic analogies between systems usually seen in physics, and~these economic systems suggest a vast amount of untapped potential in describing their~behaviour. 

In particular, in~this article, focuses on financial flows channeled into sustainability. Actors usually form a bipartite graph of investors and projects, and~the underlying dependency structure takes the form of influences between investors, and,~in general, the economic agents involved. The~dynamics influenced by this structure constitute syndicated investment in renewable energy. How is the structure of the economic interactions related to the investment dynamics? Is the behaviour universal across different energy markets (such as wind, solar, or~hydro)? Is there any observable connection between the complexity of these economic interactions, and~the rate of renewable energy investment at all? All these questions are important in climate finance, and~therefore scholars turn to the theory of complex networks to help answer~them.

This article is structured as follows. In~Section~\ref{sec:2}, we discuss the relevant background to complex networks in climate finance, and~investor networks more generally. In~Section~\ref{sec:3}, we discuss empirical evidence in different climate finance scenarios and markets. Finally, in~Section~\ref{sec:4}, we conclude with some take-away messages, and~discuss the potential for future research development in this~area.

\section{Background}\label{sec:2}
\unskip

\subsection{Econophysics and Investor~Networks}

Econophysics is a ``revolutionary reaction'' to standard economic theory that threatens to enforce a paradigm shift in thinking~\cite{rickles2011}. Usually, complex networks appear in economics within this general area. An~early and highly cited example is Mantenga's use of the graph theory to study the influence between stock prices~\cite{mantegna1999}. A~weighted complete graph is obtained from the matrix of correlation coefficients between stocks of a portfolio by considering the synchronous time evolution of
(the difference of the logarithm of) daily stock prices. For~a review of milestones and challenges in econophysics, see~\cite{kutner2019}.

Within this field, complex networks are commonplace. Remco van der Hofstad writes in his recent book on complex networks that

\begin{displayquote}
The
advent of the computer age has incited an increasing interest in the fundamental properties
of real networks. Due to the increased computational power, large data sets can now easily
be stored and investigated, and~this has had a profound impact in the empirical studies on
large networks. A~striking conclusion from this empirical work is that many real networks
share fascinating features.
\end{displayquote}

The two primary and first-studied examples of this are the scale-free degree distribution and~the small world property, known informally as \textit{{six degrees of separation}}. This universal behaviour observed in real networks has lead to the new subject of network science~\cite{hofstad2016}. As~a subdiscipline of theoretical physics, network science uses techniques and ideas from statistical physics such as random graphs, stochastic processes, combinatorics, and~wider mathematical ideas involving probability (as distinct from statistics), analysis (i.e., calculus) and dynamics (dynamical processes, particularly on networks) to reveal the structure and function of complex~systems ~\cite{aguiar2022,kartungiles2016}.


A growing trend in corporate finance is to apply centrality measures to investor networks derived from various datasets. In~Bajo et al.~\cite{bajo2020}, the value of a firm is shown to be strongly correlated with the degree of centrality of its investors in the wider US investor network (as well as with other centrality measures, in~an attempt to show the results are robust to a variety of measures). Investors are nodes, and~links form between pairs of investors when they co-invest in an equity as listed in a public US equity holding database. They write:

\begin{displayquote}
  In our sample, the~information on the equity holdings by US institutional investors allows to construct a network of relations. Stemming from the simple observation that often institutional blockholders share co-ownership
  relationships with other institutional investors, we interpret the blockholder as actor and the co-ownership link as a tie.
  \end{displayquote}

The network is then the set of actors and their ties. Fracassi~et~al.~\cite{fracassi2012} also consider centrality, showing how managers are influenced by their social peers when making corporate policy decisions, while Crane~et~al.~\cite{crane2019} show how investors acting together in cliques can amplify their voice concerning how the company is run, which strengthens governance, while weakening governance via threat of~exit.

In Dordi~et~al.~\cite{dordi2022}, ten actors are identified that can accelerate the transition away from fossil fuels using a centrality analysis of shareholder data from Bloomberg, and~the Carbon Underground 200 list of companies (200 companies that own 98\% of global oil, gas and coal reserves). The~study finds that the top ten owners of CU200 fossil fuel reserve holders are Blackrock, Vanguard, the~Government of India, State Street, the~Kingdom of Saudi Arabia, Dimensional Fund Advisors, Life insurance Corporation, Norges Bank, Fidelity Investments and Capital Group. Similarly, Galaz~et~al.~\cite{galaz2018} identify a limited set of financial actors mediating flows of capital that affect biomes of the~earth.

In Dimson~et~al.~\cite{dimson2018}, the~authors study coordinated engagements by a network of shareholders cooperating to influence firms on environmental and social
issues. They write in the conclusion that

\begin{displayquote}
 Our evidence indicates that, for~maximum effect, coordinated engagements on (ESG) issues should preferably have a credible lead investor who is well suited geographically, linguistically, culturally and socially to influencing targ\mbox{et~companies.}
  \end{displayquote}

Shareholder activist networks are studied by Yang~et~al.~\cite{yang2017}. Pension funds, special interest groups and religious organizations interact in a network of networks to influence corporate behaviour through the joint control of shares for what they perceive to be societal benefit. They show a correlation between both eigenvector and degree centrality, and~the ``efficiency of results'' obtained by the~activists.

\subsection{Nonequilibrium Statistical Physics Meets Climate~Finance}\label{subsec:windintro}

\begin{figure}[H]
  \includegraphics[width=5in]{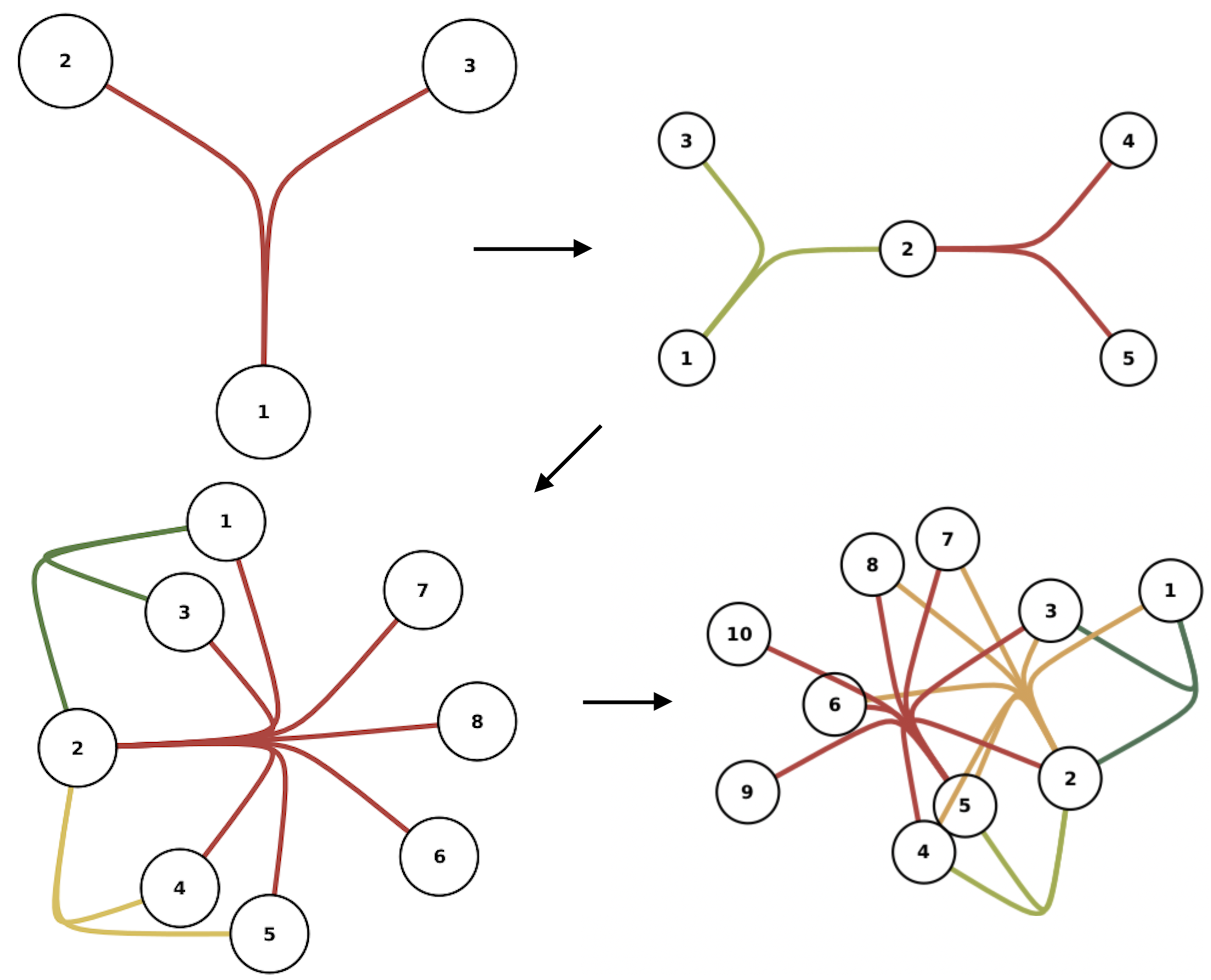}
   \caption{Hypergraph evolution. Nodes represent banks, and~hyperedges (coloured edges) represent syndicates.  New hyperedges attach to the current network nodes based on preferential attachment.}
   \label{fig:hydro4}
\end{figure}

Network evolution---see, for example Figure 
\ref{fig:hydro4}---concerns a topic within network science where growing network models, known as models of ``nonequilibrium statistical physics'', are used as null models of network growth. They attempt to explain, via simple combinatorial rules, the~large-scale universal behaviour of real networks, including their degree distribution, clustering, homology, and~anything else concerning their~structure.

An early and fundamental observation in network science is the power-law degree distribution observed in many real networks (such as citation networks, social networks, the~internet, world airline connection, etc). How does this appear? Even more important is observing it in the first place, by~comparing real networks with a null statistical model, which in the case of Barabasi and Albert is the Erdos--Renyi random graph. The~degree distribution has been known since the 1950s to follow a Poisson distribution (in the so-called thermodynamic limit where the number of nodes continues to infinity, but~keeping the expected degree constrained such that it converges to a positive constant). The~fact that random networks do not have power law degrees (also called a \textit{scale-free}
degree distribution) suggests there exist global organizing principles at play that ``fatten the tail'' or,~more formally, \textit{skew} the~degree~distribution.

\begin{figure}[H]
  \includegraphics[width=5.1in]{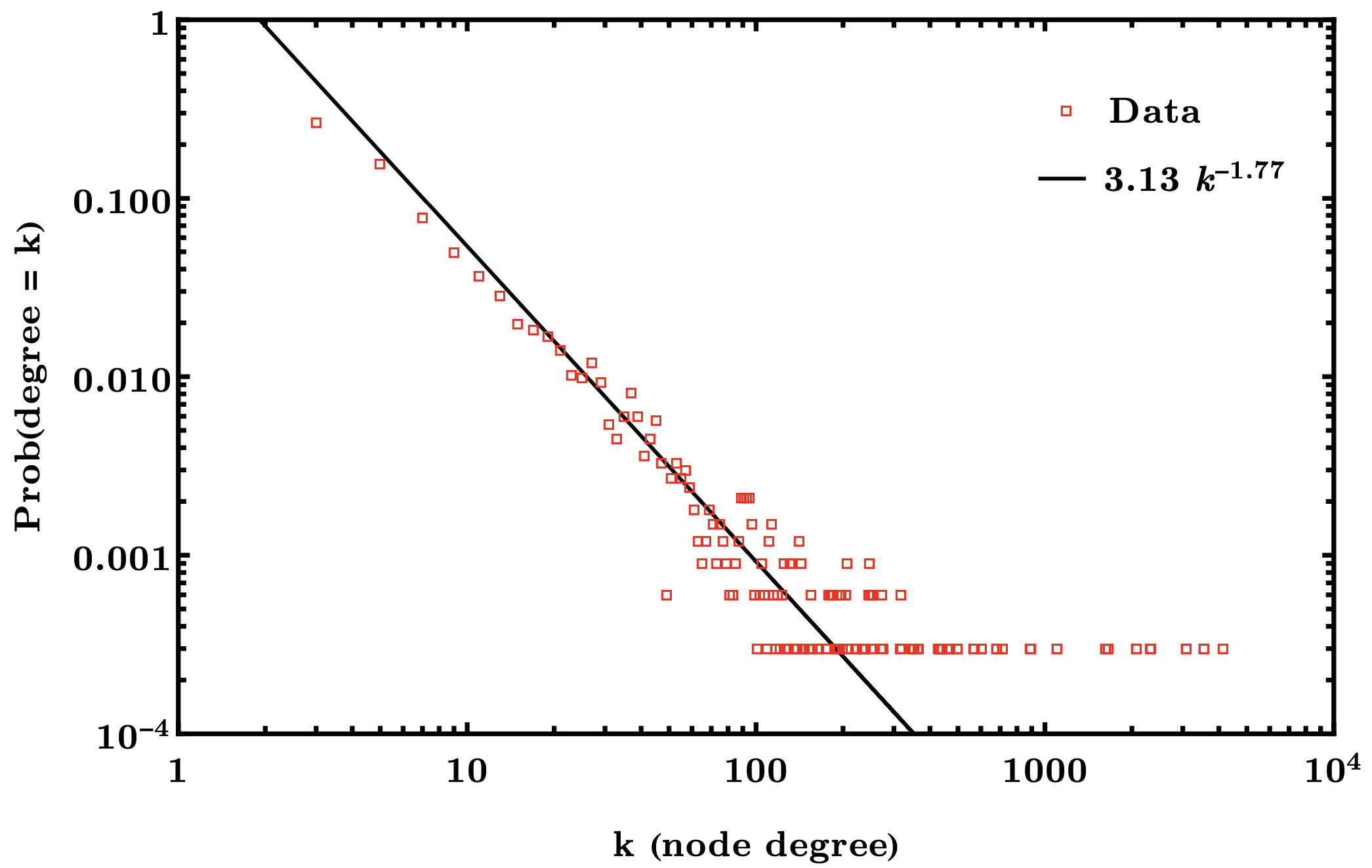}
   \caption{Degree 
     distribution for the green bond and loan network. The~red squares represent the probability a randomly selected bank in the international banking network supporting green loans and bonds is involved in $k$ bond issuances. The~non-linear model $P(k) \propto k^{-1.77}$ is the black line. This suggests a highly right-skewed degree distribution with exponent $1.77$, which occurs because the syndicates arrive in time faster than the banks, leaving a dense network where deals/banks $>1$, and~grows in time.}


 \label{fig:2}
\end{figure}

The question in finance which has only been recently explored is how this happens in financial systems such as green bond syndication networks, or~investor networks as discussed above. A~simple observation is that the degree distribution, see, e.g.,~Figure \ref{fig:2}, which is a the discrete probability mass function for the node degree (or, in~layman's terms, the~proportion of nodes with a certain degree, plotted against the degree), has an exponent $\gamma <2$. An~example hypergraph evolution model where banking syndicates of more than two parties can form is shown in Figure~\ref{fig:hydro4}.

Chung~et~al. points out that preferential attachment models of network evolution cannot explain such a large skew~\cite{chung2003}. Alternative suggested models  involve node duplication presented by Chung~et~al. in their work on biological networks. The~Pittman--Yor process is also a candidate which involves preferential attachment and~can explain degree distributions whose exponent can be lower even than unity~\cite{chung2003,allaei2006}. A~major research question is to explain the skew we observe in the green bond syndication network of Section~\ref{sec:model}. This develops early research of Rickman~et~al. in~\cite{rickman2022} and~Ameli~et~al. in~\cite{ameli2021} who also consider the effect of the fitness model of Bianconi--Barabasi~\cite{bianconi2001} using the work of Pham~et~al. on attachment functions~\cite{pham2016}.

\subsection{Investor Hubs Dominate the~Market}
The network analysis in~\cite{rickman2022} provides the first quantitative evidence of a right-skewed degree distribution. With~\textit{hubs} 
 defined in this context to be vertices of $G$ with more pairwise connections than average by one standard deviation, it is observed that  \textit{`'The domination of energy markets by a few organizations can be
driven by large incumbents achieving cost reductions through, e.g.,
economies of scale, better access to finance, or~vertical integration of
services bringing in multiple revenue streams''} (\cite[]{rickman2022}, Section 3.1
). The~authors also write that \textit{``we observe a strong positive correlation between growth of wind
markets and the level of debt hub activity''} (\cite[]{rickman2022}, Section~3.2), and~discuss this in~depth. 

\subsection{Fit Get Richer, and~Rich Get~Richer}

\textls[-19]{In Ameli et al., the~preferential attachment model is compared with the fitness model~\mbox{\cite{bianconi2001,ameli2021}}}. Instead of the standard method of considering attachment of new nodes based on the existing degree, nodes may attach to the existing lenders or sponsors based on the intrinsic fitness of nodes, as~was proposed by Bianconi and Barabasi in the Bianconi--Barabasi model introduced in 2001 to develop the theory of competition and multiscaling in evolving networks~\cite{bianconi2001}. Ameli et al. address this in the context of climate finance networks of energy efficiency~investors.

\subsection{Community~Detection}\label{sec:hydro}
Community detection is one of the largest areas of complex networks~\cite{newman2004}. The~goal is to define \textit{community} 
 in such a way that groups of financial actors are identified in a way that reveal a non-trivial structure to the larger system. Larosa et al. identify a significant home bias~\cite{lindblom2018}, writing \textit{``The investor community analysis reveals geographical investment
patterns. In~far-east Asian countries (Korea and Vietnam) the interactions between domestic investors (i.e., community density) are
more frequent compared to the rest of the world. In~India, the~financial
landscape is dominated by domestic state-owned banks, while Japan has
a strong presence over the continent through the investment made by its
second biggest bank, namely Sumitomo Mitsui Banking Corporation and
a private utility (Kansai Electric Power Co., Inc.)... Investors mainly cluster together at national and regional level confirming the existence of a “home bias” in investments''}~ (\cite[]{larosa2022}, Section 3.1).

Home bias, in~layman's terms, occurs when investors are more likely to invest in projects in their native country or region. For~obvious reasons, knowledge of the local economy and~the ability to predict the long-term prospects of a venture are a major advantage. The~community detection of Larosa et al. is the first quantitative evidence of this~effect.

Larosa~et~al. detail their methodology in \cite[]{larosa2022}, Section 2 
~using the Jaccard coefficient~\cite{murphy1996}. The~effect is, in fact, every important for the corresponding network science. Home bias leads to local clustering and~the potential emergence of network geometry, as~nearby links are favored over long range links, on~the whole. Longer range links exist, but~they connect large hubs in a way similar to the World Airlines Network~\cite{barthelemy2011}. As~such, it is important to askteh following question: To what extent does home bias lead to the emergent network structure in climate finance networks of this~type?

\subsection{Centrality~Measures}

Centrality is a measure of the importance of a node in a network. Important examples include betweenness centrality~\cite{giles2015}, where the extent to which nodes lie on multiple shortest paths in the network is considered important~\cite{kartungiles2018,kartungiles2021,knight2016}. PageRank, used by Google, is an example of providing centrality scores to websites containing a searchable keyword based on the extent to which random walks around the interconnected websites spend time on a particular~webpage.

How do we measure centrality in climate finance networks? What makes a lender node $i \in I$ or a project sponsor node $s \in S$ important to the network? Larosa~et~al. introduce a new measure based on the number of communities an investor node takes part in. They introduce the community-based centrality score (CC), writing the following: \textit{``The CC is strongly anchored to the link community structure. In~fact, well-connected 
	 investors are not just the ones with many active co-investments, but~rather those who operate in communities with high connecting power. Investors with high CC score will belong to communities capable of reaching distant groups of actors, hence spreading the available financial resources to different players. We express CC as the weighted sum of communities a node belongs to over
the X communities weighted by the average similarity between pairs of
communities''}. The~authors discuss this in further detail in \cite[]{larosa2022}, Section 2.
~Further work identifying the centrality measures important in climate finance markets is of great~interest.

\section{Empirical~Evidence}\label{sec:3}
\unskip

\subsection{Wind~Markets}

Wind makes up a significant part of renewable energy consumption around the world. For~example, in~the UK, wind makes up about a quarter of the energy contribution of the country~\cite{iea}, with~11 thousand wind turbines (14 GW onshore and 14 GW offshore) active by 2023. In~a recent journal article, \textit{The internal dynamics of fast-growing wind finance markets} \cite{rickman2022}, Rickman et al. investigate, inter alia, the claim  that preferential attachment (PA) drives the evolution of the hypergraph $G$ introduced in Section~\ref{subsec:windintro}. PA was initially introduced in a paper by Barabasi and Albert in 1999 in order to explain the emergence of the scale-free property in complex networks~\cite{barabasi1999}. The~arrival of new lenders is a discrete process of unknown temporal distribution, but~it is hypothesized in~\cite{rickman2022} that they form new links to existing equity investors with probability proportional to the attachment kernel
\begin{equation}\label{eq:le}
A_{l}(w_l) = w_{l}^{\beta_{l}},
\end{equation}
and sponsors, on~arrival, form new links to lenders with a probability proportional to the attachment kernel
\begin{equation}\label{eq:sp}
A_{s}(w_s) = w_{s}^{\beta_{s}}.
\end{equation}
With multiple lenders involved in a single project, this constitutes a hyperedge of G. When a project has recieved multiple funding sources in the form of equity or debt loans, this also constitutes a~hyperedge.

These authors do not attempt to recreate the BNEF data for the wind finance market via a random model~\cite{grimmettbook} involving preferential attachment. This aspect remains an open avenue of further research. Instead, assuming this hypothesis, the~exponents $\beta_{l}$ and $\beta_{s}$ are estimated via likelihood-based statistical methods. The~lender exponent $\beta_{l}$ of Equation~(\ref{eq:le}) and the sponsor exponent $\beta_{s}$ of Equation~(\ref{eq:sp}) are obtained via partial maximum likelihood estimation~\cite{inoue2020}. The~accuracy of the validity of the PA model is itself assessed via the likelihood ratio test of Clegg~\cite{clegg2016}; see \cite[]{rickman2022}, Section 2.5.

The authors found that the preferential attachment theory described 11 out of the 16 countries analyzed in the study. They write that \textit{debt investors (lenders) face competition for projects and past lending experience is a major determinant of who will be selected as a project partner} 
 \cite[]{rickman2022}, Section 3.3
 . The~authors claim that preferential attachment in this market is based on financial learning. Egli et al. write that \textit{``On the level of the renewable energy finance industry, investors benefitted 
 	from growing renewable energy technology (RET) markets and subsequent learning-by-doing (e.g., better risk assessment). Larger markets allowed banks to form in-house project finance teams specialized in RETs. The~knowledge and data that these teams accumulated allowed for a more accurate technology assessment. Consequently, project risks declined. For~example, as~the market had accumulated experience on historical wind speeds, investors shifted from calculating project returns on wind resource estimations with 90\% certainty''} 
 	  (\cite[]{egli2018}, Drivers of Change).

\subsection{Hydro~Markets}

Hydroelectric power (hydropower) is the largest international contributor to renewable energy production, producing more than half of the total output. Hydropower is particularly popular in developing countries, and~thus plays an important role in the UN sustainability goals~\cite{iea}. Larosa et al. write that \textit{``financing hydropower projects requires investors to pay large upfront capital and lock in 
	 their capital for decades (hydro projects can last for 100 years), while also bearing high
investment risks''}. 
 With~this in mind, the~hydroelectric project financing landscape is addressed by Larosa et al. in \textit{``Finding the right partners? Examining inequalities in the global investment
  landscape of hydropower''} 
   \cite{larosa2022}. Given the unique aspect of intercontinental development at work, internationally diverse financial actors need to be assembled. The~focus is therefore on centrality and community detection rather than network evolution; see Figure 3 in \cite{larosa2022}.





Financing hydropower projects necessitates substantial upfront capital investment, with~funds tied up for extended periods, often spanning a century due to the long lifespan of such projects. Indeed, the~construction of a large hydropower dam typically exceeds a billion dollars, demanding patient capital and enduring the natural investment cycle. Thus, an~intricate network of diverse investors, and~effective capital distribution becomes essential for hydropower assets. The~focus is therefore on centrality and community detection rather than network~evolution.

\subsection{Energy Efficiency~Markets}


Energy Efficiency Technologies are interventions that reduces energy consumption, such as using light-emitting diodes (LED) in place of conventional filament bulbs. This technology has a major place in modern science due primarily to its efficiency. The~Nobel Prize in 2014 was awarded for the blue LED as it enables white light and a~more universal employment of energy efficiency with the climate in mind~\cite{led1}. Ameli~et~al. write that \textit{``investments in energy efficiency (EE) are particularly crucial to reduce the energy demand for a growing world economy and are listed as core measures for sustainable recovery plans''} \cite{ameli2021}.

As with the research in wind markets, Ameli~et~al. focus on the theme of preferential attachment in a bipartite graph of investors and energy efficiency projects. Applying ideas from Pham~et~al. in their recent work concerning the joint estimation of preferential attachment and node fitness in growing complex networks, the~authors look at how influential the intrinsic fitness of a node to acquire links is compared with its degree-based link acquisition (i.e., simple preferential attachment compared with a fitness-based network evolution model~\cite{bianconi2001}).

They \textit{``empirically estimate the preferential attachment (PA) function and node fitnesses from observed network data''} \cite{ameli2021}. The~authors suggest that there is a balance between preferential attachment and node fitness determining the evolution of their network, writing the following: \textit{``Following Pham~et~al. approach, we measure the respective influences of the preferential attachment and the fitness models''} \cite{ameli2021, pham2016}. The~PAFit method is discussed by Pham: \textit{``Our main contributions are twofold. The~first contribution is a statistical method called PAFit to simultaneously estimate the PA and node fitness functions without imposing any assumptions on their functional forms. To~the best of our knowledge, PAFit is the first ever method in the literature that can do so''}.

\begin{figure}[H]
  \includegraphics[width=4.5in]{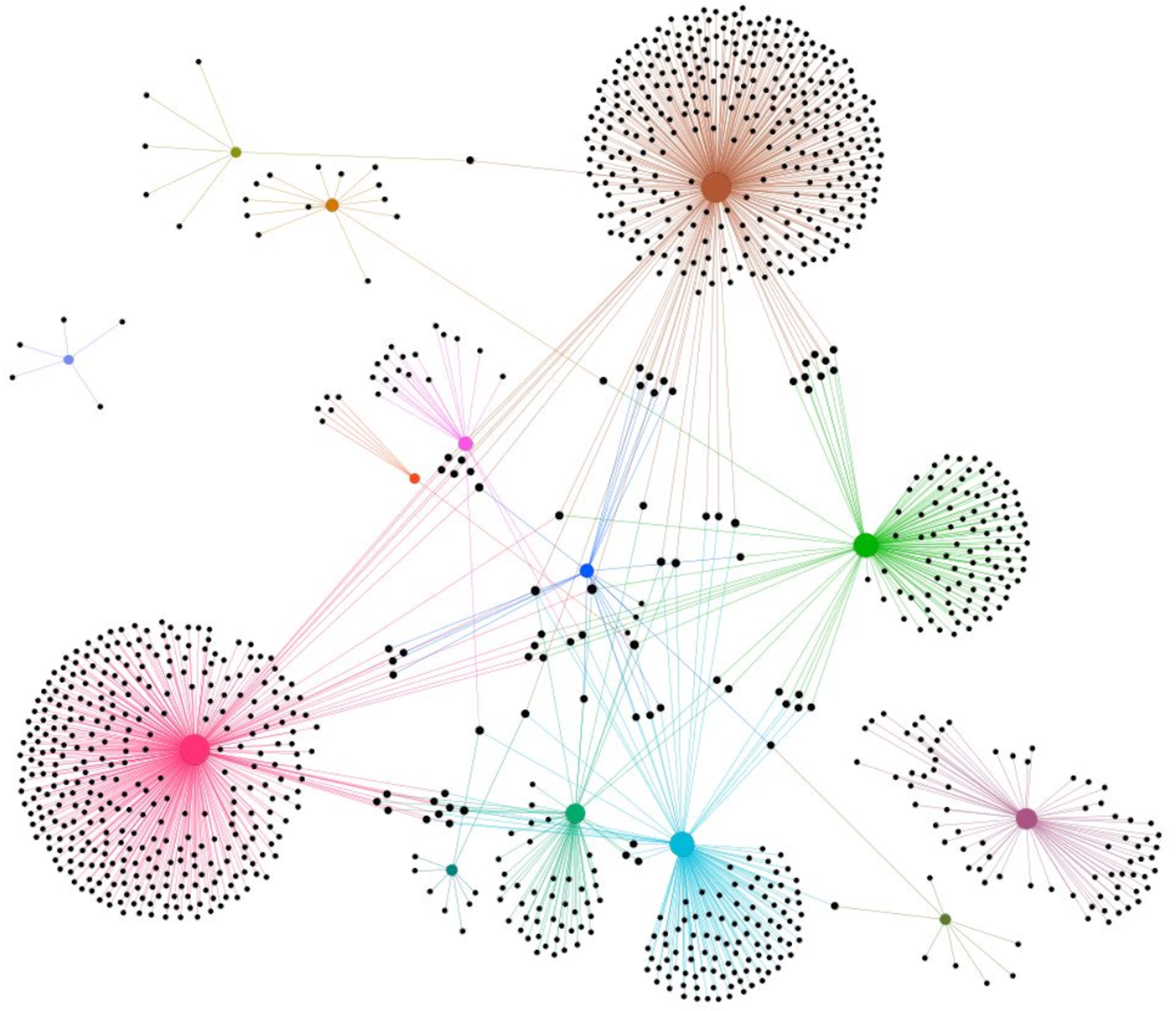}
   \caption{Aggregated 
 network of financial actors involved in energy efficiency financing (2000--2017) in different sectors, {taken from}~\cite{ameli2021}.
~Nodes are investors, and edges are financial interactions between them, with the following key: pink from state-owned utility, brown from investor-owned utilities, light blue from manufacturing and services, green from the governmental sector, dark purple from an energy cooperative, light-green research and the university sector,  blue from institutional investors, orange construction and real estate, turquoise diversified, deep green chemicals and steel, green-brown food, bright red retail, light purple defence, and bright purple the remaining uncatagorised areas,
. }
 \label{fig:3}
\end{figure}

Given this approach to the theory of network evolution, the~authors draw the conclusion that \textit{``...this suggests that the ‘rich get richer’ mechanism becomes weaker when the ‘fit get richer’ effect is considered, showing that to some extent technology’s ability to attract new investment is explained by its fitness''}. They also discuss the snapshot overtime of the network 
(see Figure~\ref{fig:3}), observing the total number of investments that different types of investors (e.g., from the utilities sector) have made (\cite[]{ameli2021}: evolution, dynamics and growth of the energy efficiency network).
\vspace{-1pt}

\subsection{Green Bonds, Loans, and~Networks of Underwriter~Syndicates} \label{sec:model}

Green bonds, loans and debt securities are designated to finance environmentally friendly projects. This may take the form of renewable energy infrastructure, such as a wind farm, or~ refurbishment of real estate to make it more sustainable. Whatever project requires funding, the~project managers approach a bank looking for funding. They attempt to acquire investment by selling green bonds to investors~\cite{rickman2022}. These are underwriting, i.e.,~insured by a banking syndicate who buys all the bonds and resells them to investors for profit,~therefore taking on the risk in case the project collapses. This guarantees returns for the holders of the bonds (these may be private customers, such as pension funds, or~individuals purchasing online using their own funds) \cite{wikipf}.



\begin{figure}[H]
  
   \includegraphics[width=5.4in]{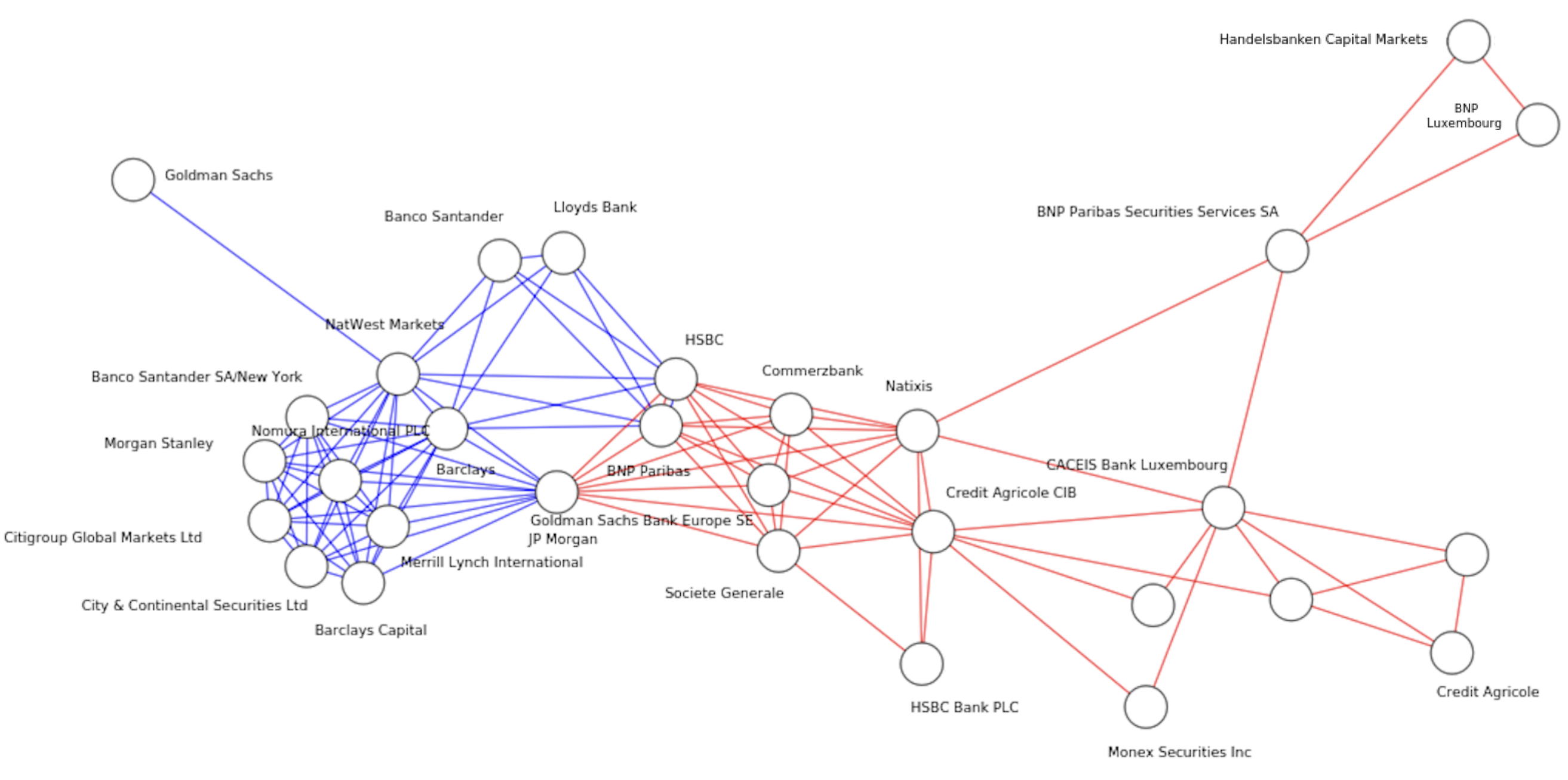}
\caption{A sample of the multilayer network of banks (nodes) underwriting green bonds and loans in two countires, UK (blue links), and France (red links). Goldman Sachs, BNP Paribas, and HSBC connect the layers, serving as international actors which unite layers more often than local banks.}
 \label{fig:5}
\end{figure}

One can build a complex network---see Figure~\ref{fig:5}---from~transaction data in the following way. The~modeling consists~of

\begin{enumerate}
\item A hypergraph $G(V,E)$ where $V$ is the vertex set and $E$ is the edge set, with~$|V|=n|$ and $|E|=m$, and~each edge $e$ is simply a subset of $V$; see \cite[]{chen1996}, Introduction.
 
\item The vertices, which represent~banks.

\item The hyperedges (i.e., higher-order edges representing groups of investors and an investment rather than simply pairs of investors). which represent project financing by the corresponding banking syndicate. The~amount of money invested is large enough in many cases to require large syndicates of banks to underwrite the risk.
\end{enumerate}

See also Berge~\cite{berge1989} and Beckenbach~\cite{beckenbach2019} for a discussion of bipartite hypergraphs. Note that there is also a potential to view this as a simplicial complex~\cite{kartungiles2019}. Higher-order network models of a banking network is an interesting area of further research in this~area.



 \section{Final Words and Open Avenues of~Research}\label{sec:4}

 The area is still developing, but~has immense potential. Understanding the interesting aspects of complex networks which present particular application in banking networks funding climate initiatives allows policy makers intervention and influence on climate funding in a postive way for~society.

The ways in which different market places, different sectors such as finance, technology or utilities, or~different geographic regions lead to different network structure, or~the ways in which it is universal, is still not well understood. 
 Community detection needs further work, for~example, by developing network embedding techniques that incorporate financial metrics (e.g., weighted edges representing money mobilized in a deal). As~we disucssed, higher-order network models of a banking network is an interesting area of further~research.

Two recent articles have addressed preferential attachment as the main driver of network evolution. It remains an open question as to whether random models of climate finance hypergraphs which evolve based on ''rich get richer'' or~''fit get richer'' models are able to reproduce the data of BNEF in a sophisticated way. This would present an important connection between statistical physics and climate finance, and~allow further insights into how these networks evolve and develop. How to then encourage the transition to green energy based on this detailed understanding is a difficult and multi-disciplinary task, but~well founded on the excellent descriptive analysis that can be provided by these early works in complex networks  Further work on network evolution is critical to understand the mechanisms that generate the highly skewed degree distributions observed in banking syndicate networks. We look forward to a future review concerning research developing these ideas, and~to the corresponding new insights into climate finance as we track the critically important goals of the Paris~agreement.

We hope that in the future, the~link between policy and network structure can be addressed, as well as the ways in which this structure leads to better and more sustainable green growth. This is a major challenge which we hope the networks community can begin to address to~provide a remarkable example of physics in~society.

\vspace{6pt}

\authorcontributions{Writing---original draft, A.P.K.-G. and N.A. All authors have read and agreed to the published version of the manuscript. 
}

\funding{Both authors acknowledge support from the European Research Council (ERC) under the European Union’s Horizon 2020 research and innovation programme (grant agreement No 802891).
}



\dataavailability{Bloomberg data used for any banking syndicate informaton (not published elsewhere) is proprietary and shareable on request with the corresponding author.}


\acknowledgments{We thank Denitsa Angelova, Ginestra Bianconi, Claudia Brown,  Max Falkenberg, Michael Grubb, Ben Hinder, Francessca Larosa, Figo Lau, Sumit Kothari, and Jamie Rickman, for many helpful discussion.}

\conflictsofinterest{The authors declare no conflict of interest. 
}

\begin{adjustwidth}{-\extralength}{0cm}

\reftitle{References}

\PublishersNote{}
\end{adjustwidth}
\end{document}